\documentclass[a4paper,11pt]{article}

\usepackage{graphicx}
\usepackage{float}
\usepackage{subfig}

\usepackage{jinstpub} 
\usepackage{lineno}
\title{\boldmath Performance of AC-LGAD strip sensor designed for the DarkSHINE experiment}

\author[a,b,c,1]{Kang Liu,\note{co-first author.}}
\author[d,e,f,1]{Mengzhao Li,}
\author[a,b,c]{Junhua Zhang,}
\author[d,f]{Weiyi Sun,}
\author[d,e]{Yunyun Fan,}
\author[d,e]{Zhijun Liang,}
\author[a,b,c,2]{Yufeng Wang,\note{co-corresponding author.}}
\author[d,e,2]{Mei Zhao,}
\author[a,b,c,2]{Kun Liu}

\affiliation[a]{Tsung-Dao Lee Institute, Shanghai Jiao Tong University, Shanghai 201210, China;}
\affiliation[b]{ Institute of Nuclear and Particle Physics, School of Physics and Astronomy, Shanghai Jiao Tong University, Shanghai 200240, China;}
\affiliation[c]{Key Laboratory for Particle Astrophysics and Cosmology (MOE), Shanghai Key Laboratory for
Particle Physics and Cosmology (SKLPPC), Shanghai 200240, China;}
\affiliation[d]{Institute of High Energy Physics, Chinese Academy of Sciences, Beijing 100049, China;}
\affiliation[e]{State Key Laboratory of Particle Detection and Electronics, Institute of High Energy Physics, Chinese Academy of Sciences, Beijing 100049, China;}
\affiliation[f]{University of Chinese Academy of Sciences, Beijing 100049, China}
\emailAdd{kun.liu@sjtu.edu.cn}

\abstract{AC-coupled Low Gain Avalanche Detector (AC-LGAD) is a new precise detector technology developed in recent years. Based on the standard Low Gain Avalanche Detector (LGAD) technology, AC-LGAD sensors can provide excellent timing performance and spatial resolution. This paper presents the design and performance of several prototype AC-LGAD strip sensors for the DarkSHINE tracking system, as well as the electrical characteristics and spatial resolution of the prototype sensors from two batches of wafers with different $n^+$ dose.The range of spatial resolutions of 6.5~$\mathrm{\mu m}$ $\sim$ 8.2~$\mathrm{\mu m}$ and 8.8~$\mathrm{\mu m}$ $\sim$ 12.3~$\mathrm{\mu m}$ are achieved by the AC-LGAD sensors with 100$\mu m$ pitch size.}


\arxivnumber{xxxx.xxxx} % only if you have one

\begin{document}

\maketitle

\flushbottom

%------------------------------------------------------------
%------------------------------------------------------------
%------------------------------------------------------------
%------------------------------------------------------------
\section{Introduction}
\label{sec:Introduction}

DarkSHINE experiment\cite{DarkSHINE} is a newly proposed electron-on-fixed-target experiment searching for dark photon produced via electron and nucleon interaction. The dark photon then decays to a pair of dark matter candidates, which is called invisible decay\cite{darkphoton}. Dark matter pairs escape detection as missing momentum and missing energy, also resulting in lower momentum and larger recoil angle of the recoil electron. The missing momentum signature is used to identify signal from various Standard Model background processes. Therefore, reconstruction of position and momentum of the incident and recoil electrons is crucial for this experiment. 

\begin{figure}
\centering 
\subfloat[DarkSHINE experiment.]{\includegraphics[width=.45\textwidth]{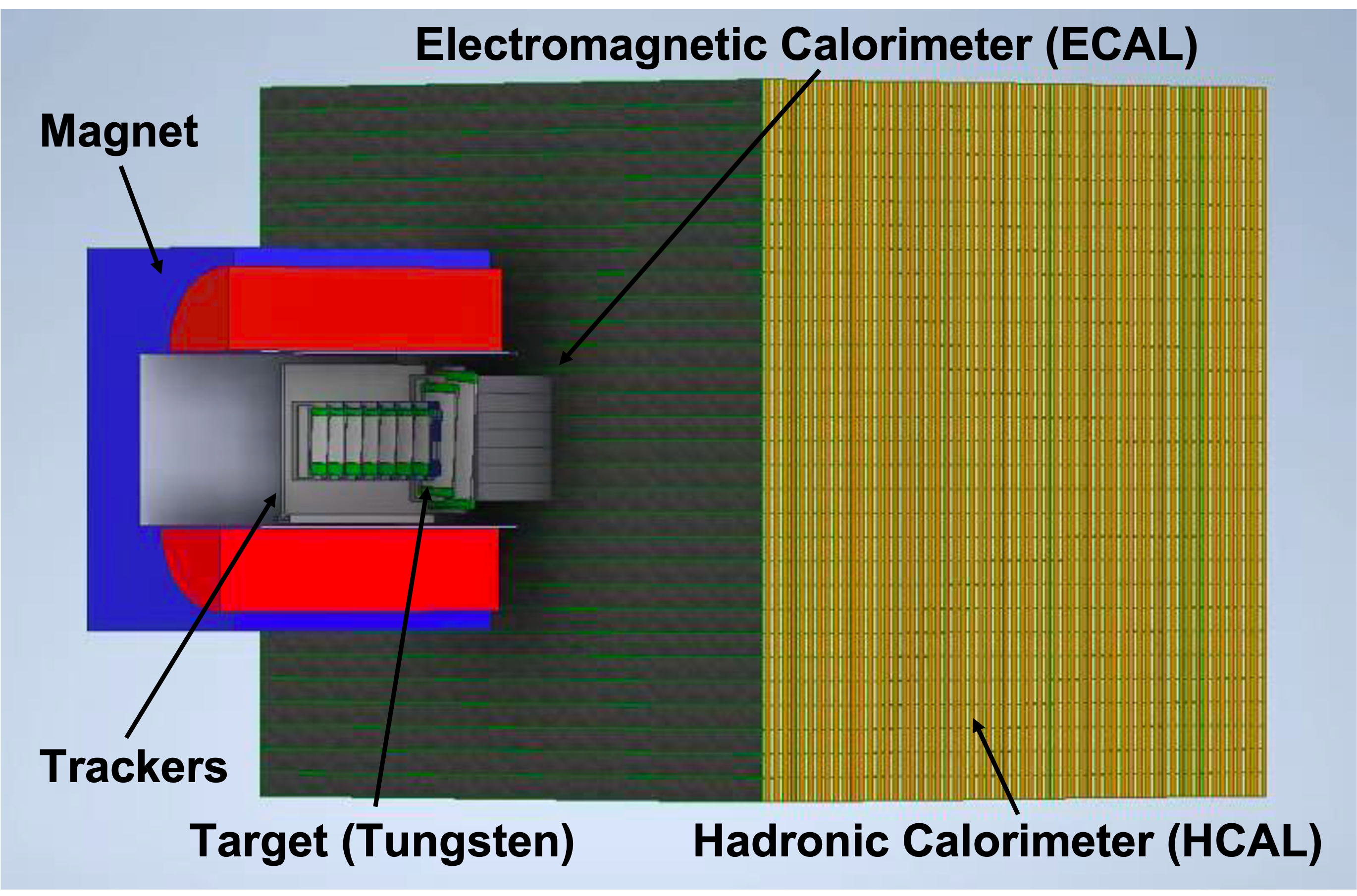}}
\qquad
\subfloat[DarkSHINE tracking system.]{\includegraphics[width=.45\textwidth]{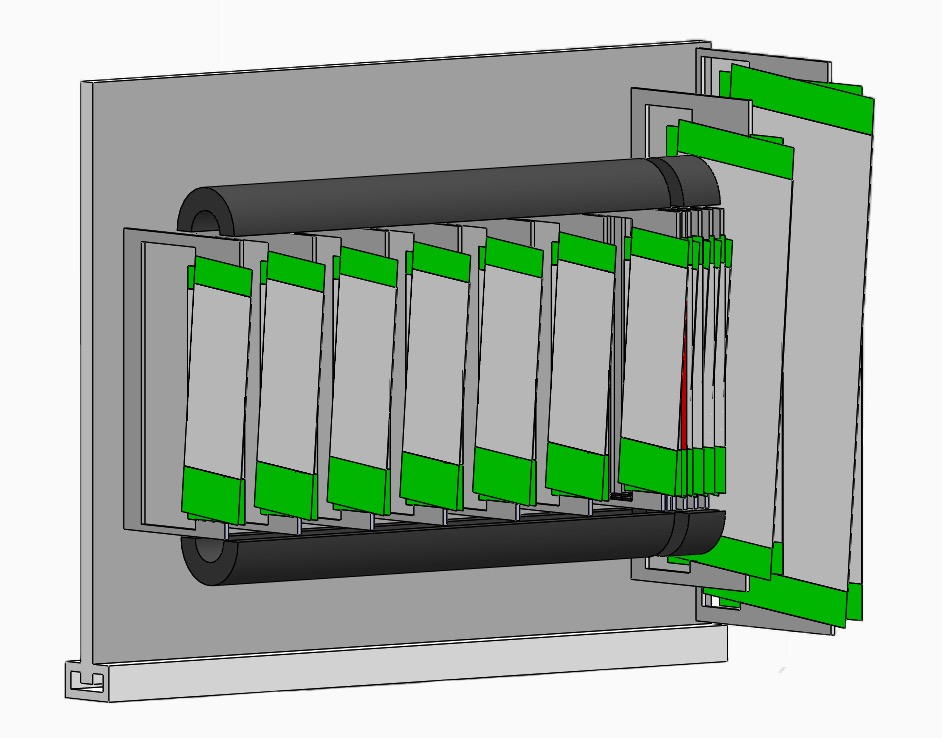}}
\caption{\label{fig:DarkSHINEexperiment} (a) Detector sketch of the DarkSHINE experiment\cite{DarkSHINE}. Along the electron incident direction from left to right in the picture, the red material with a blue brace is the dipole magnet. The tagging tracker is placed at the center of it. The recoil tracker is located at the edge of the magnet. The target is caught in the middle. ECAL is placed after the recoil tracker, followed by HCAL. (b) Sketch of the DarkSHINE tracking system. From left to right:  tagging tracker (with seven layers of tracking module), the tungsten target, and the recoil tracker (with six layers of tracking module). For each layer, two strip sensors are shown in the sketch, placed at a small angle (100 mrad).}
\end{figure}

Figure~\ref{fig:DarkSHINEexperiment}-(a) shows the sub-detector systems of DarkSHINE experiment. A tracking system is placed in a downward magnetic field of around 1.5~T which is provided by a superconducting magnet system. The magnetic filed direction is defined as the $y$-direction and the electron beam direction as the $z$-direction, hence the electron will be deflected in the $x$-direction perpendicular to the magnetic field. As shown in Figure~\ref{fig:DarkSHINEexperiment}-(b), the DarkSHINE tracking system consists of seven layers of tagging modules and six layers of recoil modules, while a tungsten target with $0.1 \mathrm{X_0}$ decay length is placed in between. Each layer of tracking module consists of two silicon sensors with length of at least 20 mm, placed at a small angle (100~mrad). The sensors are expected to be as thin as possible, in order to avoid multi-track events caused by the interaction between charged particles and the nucleus of the detector material. The designed position (angle) resolution of the tracking system is better than 10 $\mathrm{\mu m}$ (0.1\%) at the direction of the electron deflection. To achieve that, several small prototype sensors are designed with technology of AC-coupled Low-Gain Avalanche Detectors (AC-LGADs, also called Resistive Silicon Detectors).

\begin{figure}[htbp]
\centering 
\subfloat[LGAD sensor.]{\includegraphics[width=.45\textwidth]{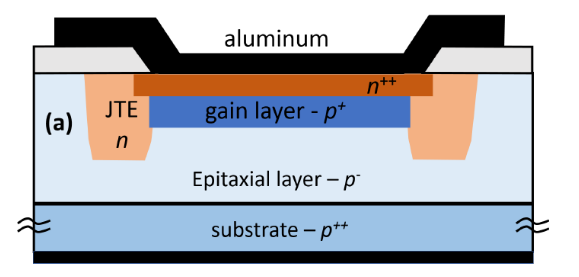}}
\qquad
\subfloat[AC-LGAD sensor.]{\includegraphics[width=.45\textwidth]{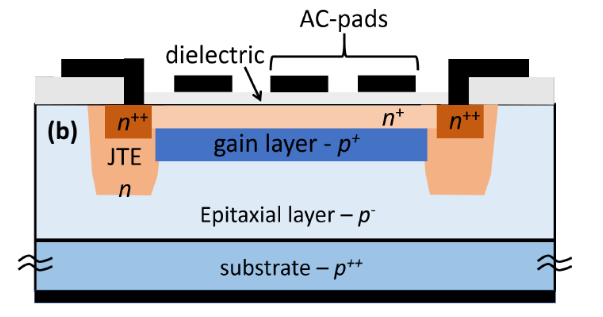}}
\caption{\label{fig:Introduction} (a) Sketch of a section of a single-pad standard LGAD and (b) a section of a segmented AC-LGAD sensor\cite{Giacomini:2019kqz}. The sketches are not to actual scale.}
\end{figure}

%--------------------------------
The Low-Gain Avalanche Detector (LGAD)\cite{HFW.Sadrozinski} has been developed in recent years, with a novel precise detector technology initially proposed and designed for the precise timing measurements of the High Granularity Timing Detector of ATLAS\cite{ATLAS} and the Endcap Timing Layer of CMS\cite{CMS} for the High-Luminosity Large Hadron Collider. Figure~\ref{fig:Introduction}-(a) shows a sketch of a LGAD sensor. The LGAD sensors are fabricated on high-resistivity $p$-type substrates with thickness of about 50~$\mathrm{\mu m}$. Based on traditional $n$-in-$p$ silicon sensor, the LGAD sensor has an additional highly-doped $p^+$ region (namely ``gain layer'') under the parallel $n$-$p$ junction. When a bias voltage is applied across the sensor, this $p^+$ layer will be depleted and create a strong local electric field, therefore introducing an internal gain.

In order to achieve better spatial resolution while maintaining similar gain and fast timing performance, the AC-coupled LGAD (AC-LGAD) technology has been brought up where the signal are capacitively induced and shared among the metal AC-pads. Figure~\ref{fig:Introduction}-(b) shows a sketch of an AC-coupled LGAD sensor. The AC-pads of the sensor for signal readout are placed on a thin dielectric layer which is grown over the $n^+$ layer of the sensor. The AC-LGADs has much less doped $n^+$ layer comparing to the $n^{++}$ layer in the standard LGADs. It results in an increased inter-pad resistance\cite{Giacomini:2019kqz}. A highly-doped $n^{++}$ implant is still preserved at the edge of the active area of the sensor, in order to have DC-connection for electron current draining. The AC-LGAD design can easily adapt to any detector geometry since the segmentation can be achieved by the AC-pads of any shape.

\begin{figure}[htbp]
\centering 
\subfloat[Strip sensor design.]{\includegraphics[width=.46\textwidth]{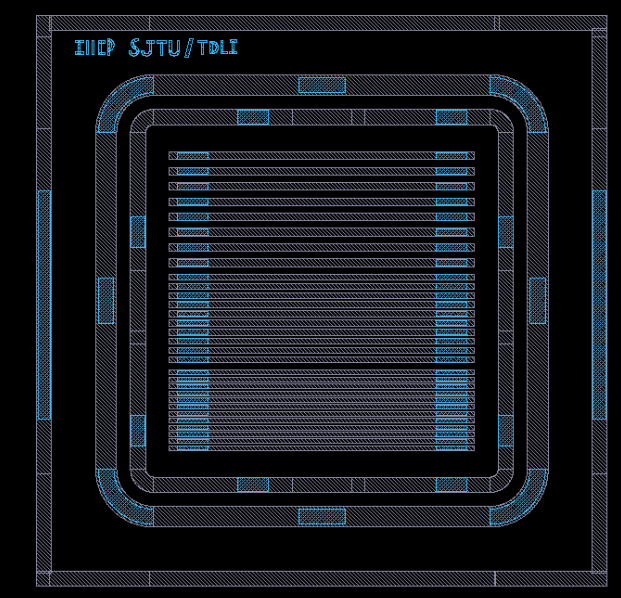}}
\qquad
\subfloat[Strip sensor picture.]{\includegraphics[width=.44\textwidth]{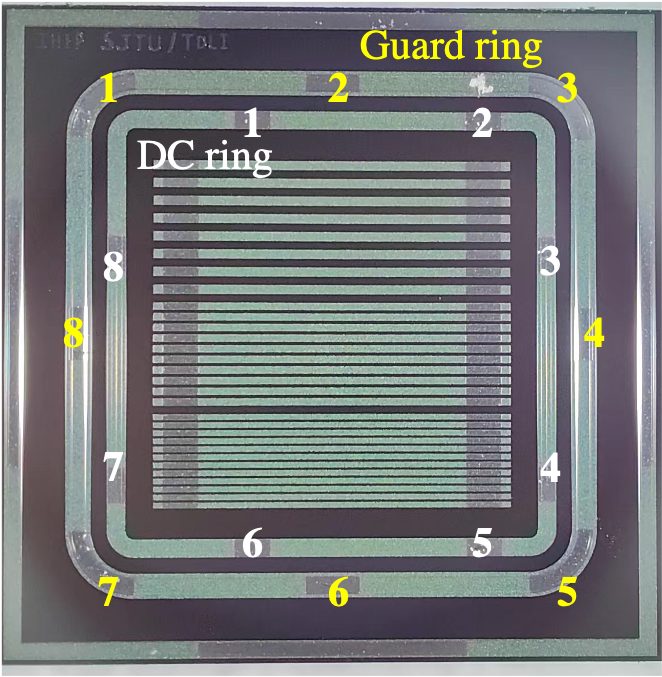}}
\caption{\label{fig:sensor} (a) Design of the AC-LGAD strip sensor for DarkSHINE. (b) picture of the silicon strip sensor under microscope.}
\end{figure}

Two types of strip sensors with pitch (strip) size of 100~(50)~$\mathrm{\mu m}$, 60~(40)~$\mathrm{\mu m}$, and 45~(30)~$\mathrm{\mu m}$ are designed for the DarkSHINE experiment. Two batches of wafers of AC-LAGD strip sensor prototypes have been produced by the Institute of Microelectronics, which are based on the AC-LAGD technology designed by the Institute of High Energy Physics (IHEP), CAS. They are named as wafer-11 and wafer-12 sensor in the following context. The $n^+$ doses of these two wafers are 0.01~P and 10~P, where P is the unit of phosphorus dose defined for the AC-LGADs\cite{N+dosePaper}. Smaller dose leads to higher resistance of the sensor and therefore smaller leakage current. Thus it is expected to have better performance on the position reconstruction and spatial resolution. Figure~\ref{fig:sensor} shows the design and a picture of a fabricated AC-LGAD strip sensor prototype (size of $1\times 1~\mathrm{mm^2}$). There are three rings which are (from outside to inside) the outer ring, the Guard Ring (GR), and the DC ring. The sensors have a 50~$\mathrm{\mu m}$ epitaxial layer and a 725~$\mathrm{\mu m}$ substrate.

This paper presents performance of the AC-LGAD prototype strip sensors with pitch (strip) size of 100 (50) $\mu m$. Section~\ref{sec:AC-LGAD sensor performance} introduces the electrical characteristics of the prototype sensors. Section~\ref{sec:Position reconstruction and spatial resolution} discusses position reconstruction and the spatial resolution measurements. 

%------------------------------------------------------------
%------------------------------------------------------------
\section{AC-LGAD sensor I-V and C-V performance test}
\label{sec:AC-LGAD sensor performance}

\subsection{Measurement setup}
\label{sec:Measurement setup 1}

\begin{figure}[htbp]
\centering 
\subfloat[I-V test.]{\includegraphics[width=.45\textwidth]{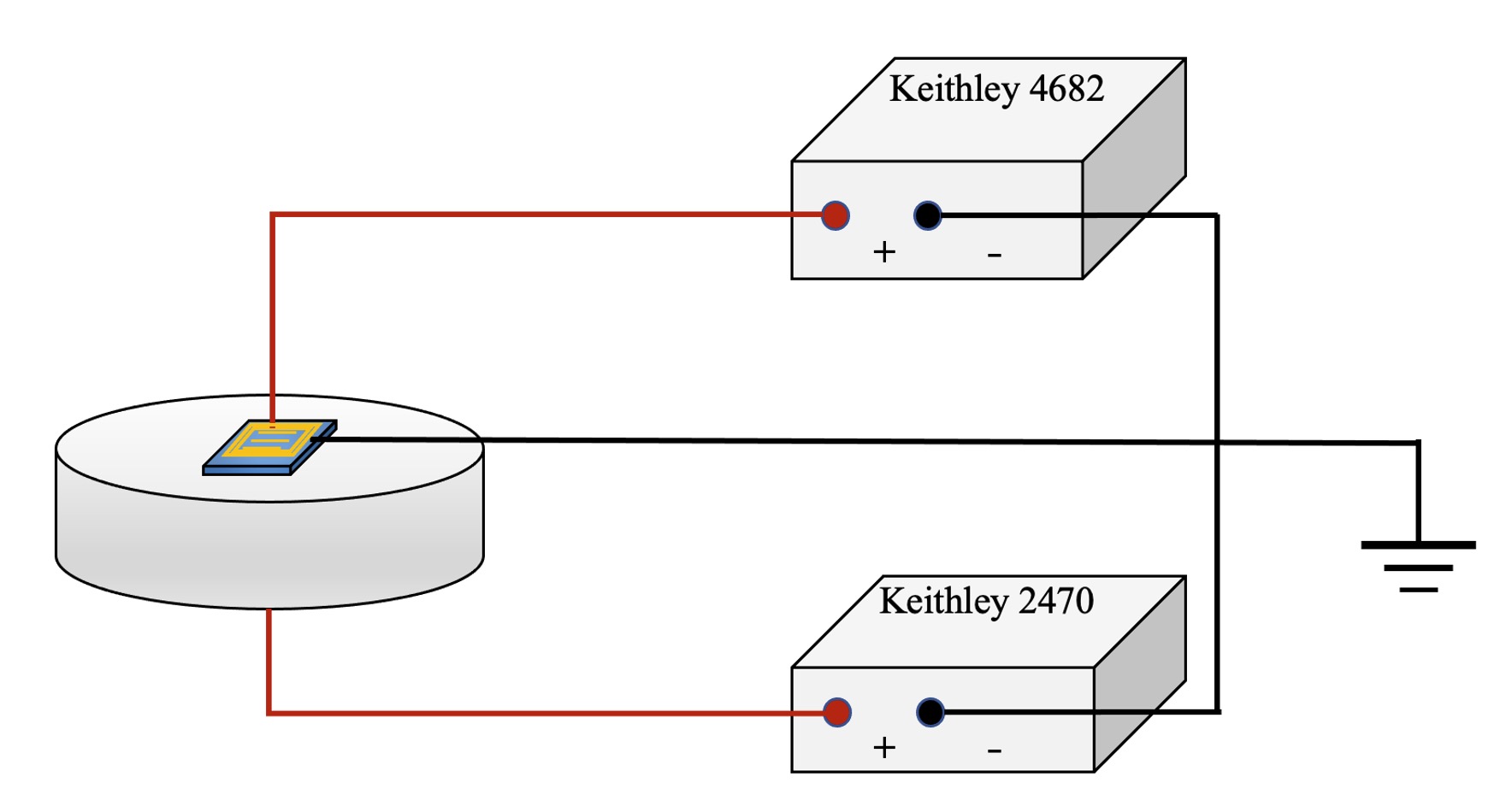}}
\qquad
\subfloat[C-V test.]{\includegraphics[width=.45\textwidth]{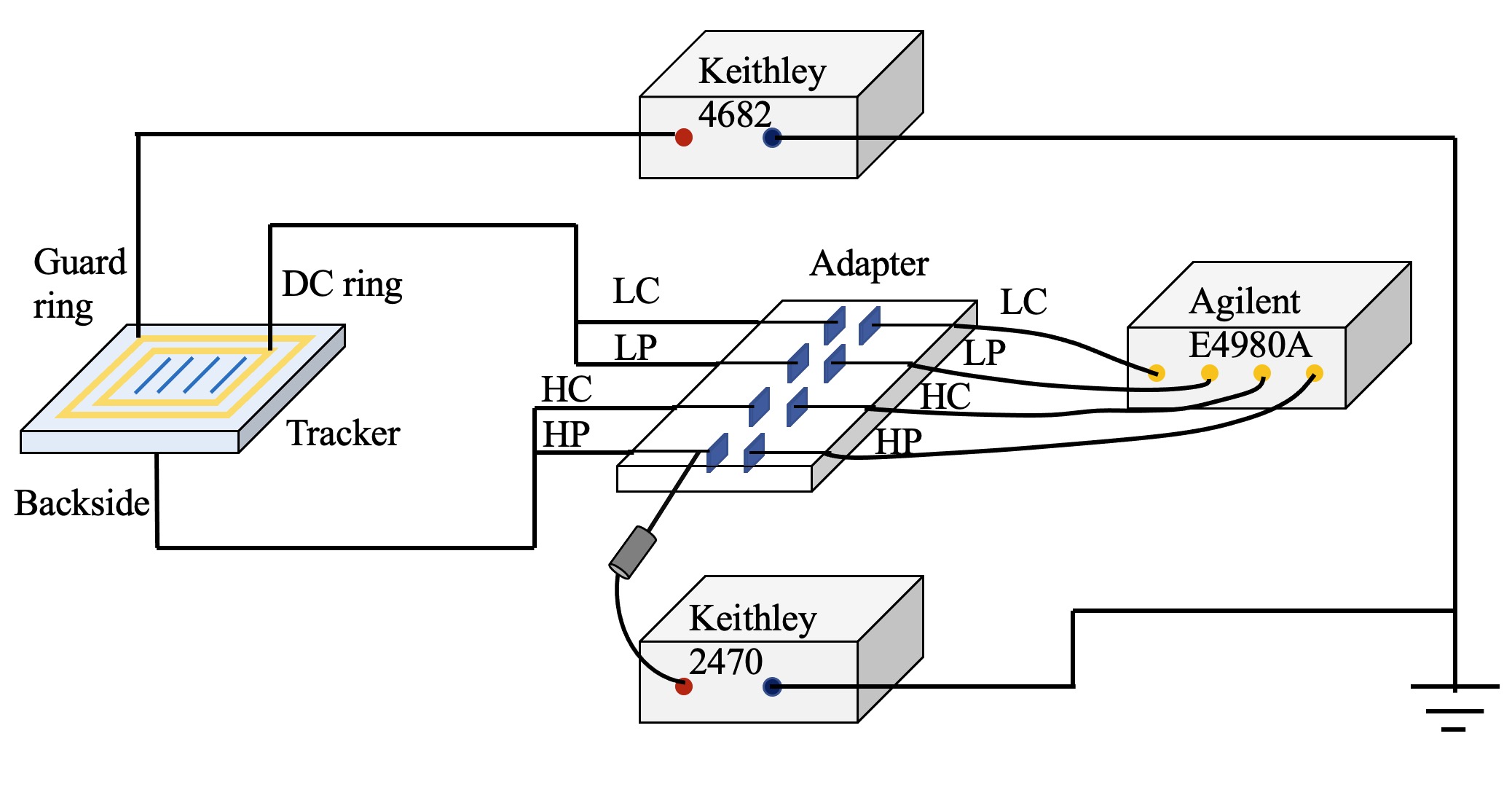}}
\caption{\label{fig:Wire} Wiring setup for the (a) I-V tests; (b) C-V tests.}
\end{figure}

The sensors I-V and C-V tests are carried out in the tracking lab at Shanghai Jiao Tong University. The wiring setups for I-V and C-V measurements are illustrated in Figure~\ref{fig:Wire}. A four-channel (LC, LP, HC, HP) HV adapter is connected in parallel between the sensor and the LCR meter to limit the voltage within a safe range. In  C-V measurement, the LCR meter works on the ``Cp-Rp'' mode at a frequency of 10kHz according to the recommendation of RD50 \cite{RD50Recommendation}. The measurements are carried out in darkness to avoid additional light current.

%------------------------------------------------------------
\subsection{Current-voltage measurement}
\label{sec:I-V test}

\begin{figure}[htbp]
\centering 
\subfloat[]{\includegraphics[width=.45\textwidth]{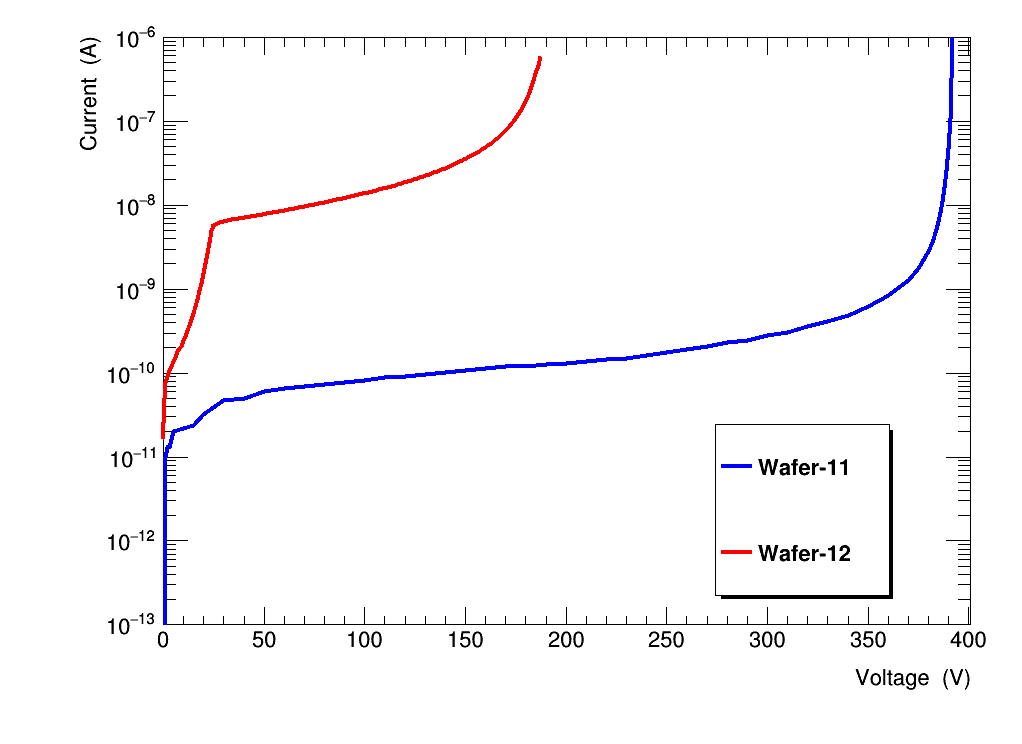}}
\qquad
\subfloat[]{\includegraphics[width=.45\textwidth]{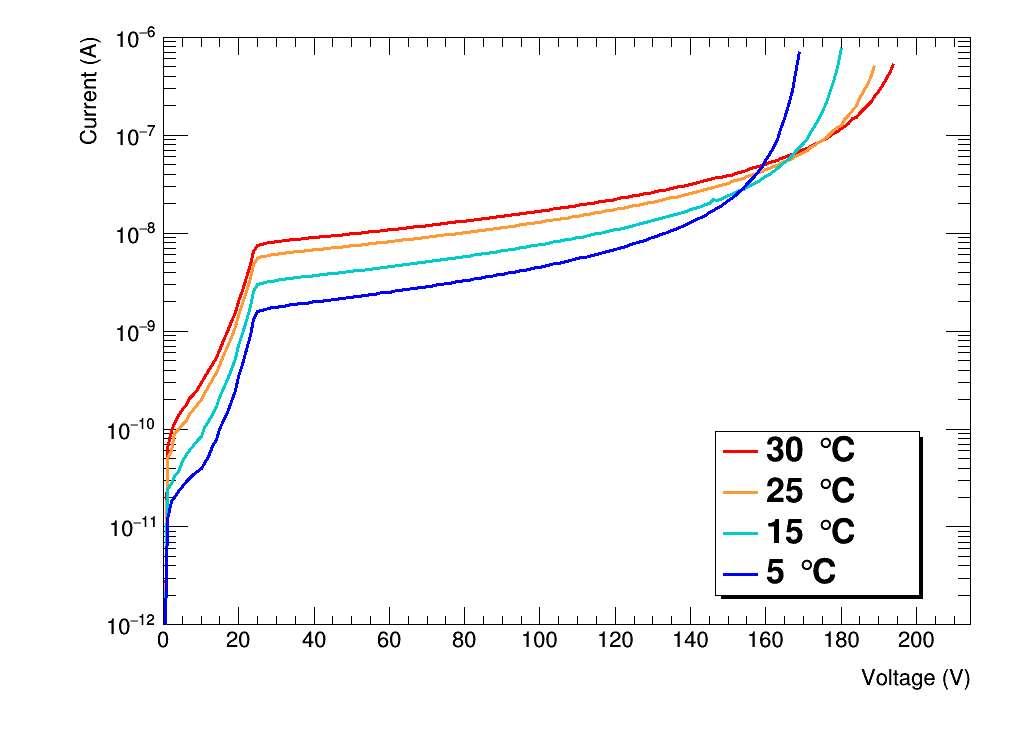}}
\caption{\label{fig:IVt_emp_uni} (a) Current-Voltage performance of two AC-LGAD sensors from wafer-11 (blue) and wafer-12 (red). (b)Current-Voltage temperature dependency of wafer-12 sensors.}
\end{figure}

The measured I-V performance is shown in Figure~\ref{fig:IVt_emp_uni}. The breakdown voltage~($V_{BD}$) of wafer-11 (wafer-12) sensor, which is defined as the reverse bias voltage applied when the leakage current reaches 500 nA, is around 380~V (185~V) at room temperature (~25$^{\circ}$C). Wafer-11 has higher $V_{BD}$ than wafer-12 due to higher resistance. Figure~\ref{fig:IVt_emp_uni}-(b) shows the I-V curves measured at different temperatures: 5$^{\circ}$C, 15$^{\circ}$C, 25$^{\circ}$C and 30$^{\circ}$C. As temperature increases, the current and $V_{\mathrm{BD}}$ also increases due to thermal motion of the electron-hole pairs. This result suggests that working voltage of the wafer-11 (wafer-12) sensors should be set below 350~V (150~V).

\begin{figure}[htbp]
\centering 
\subfloat[]{\includegraphics[width=.45\textwidth]{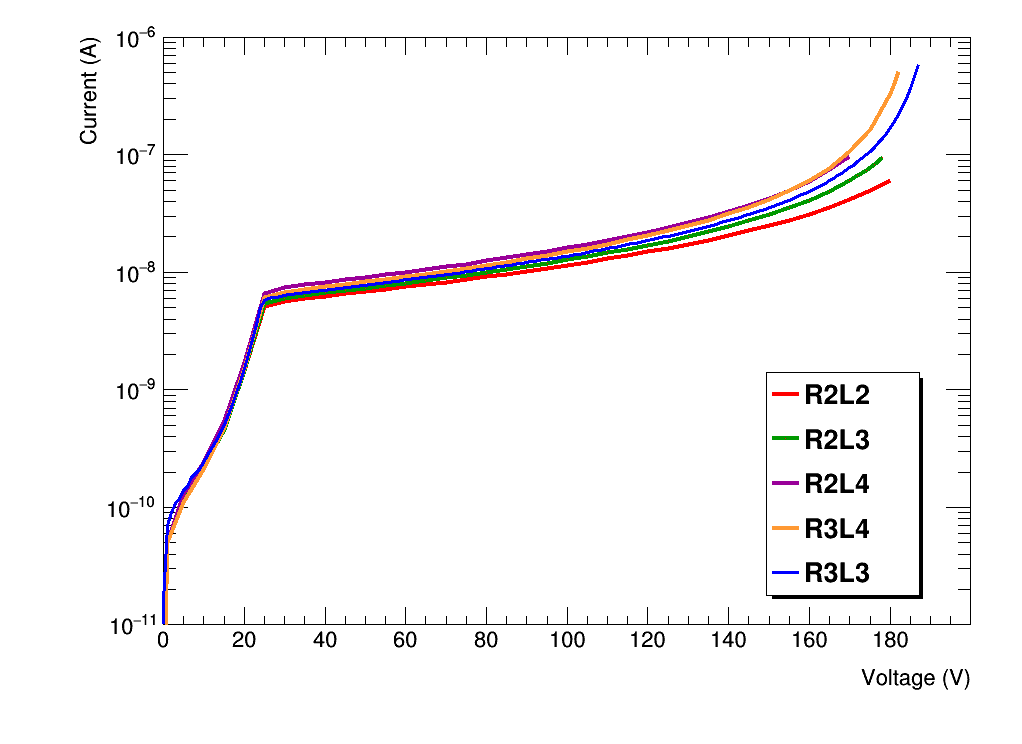}}
\qquad
\subfloat[]{\includegraphics[width=.45\textwidth]{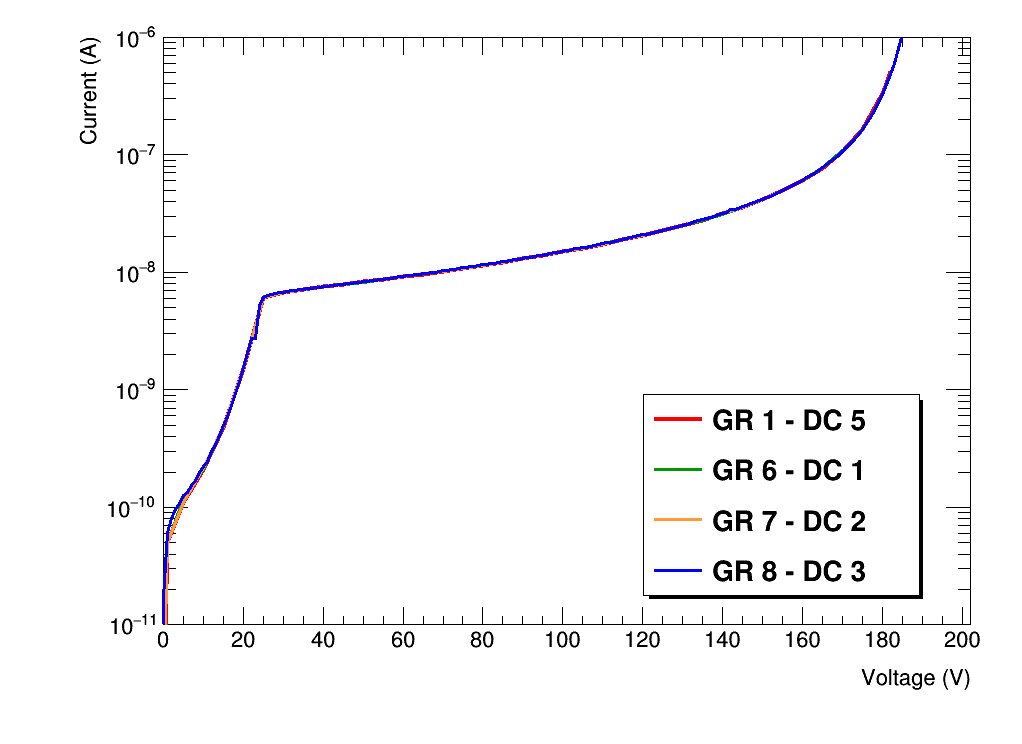}}
\caption{\label{fig:IV_all} Current-Voltage performance of AC-LGAD sensors from wafer-12. (a) Curves with different color represents sensor from different position on the wafer: R indicates the row and L indicates the column of the sensor. (b) probe needles on different position of the GR and the DC ring, as marked in Figure~\ref{fig:sensor}.}
\end{figure}

Figure~\ref{fig:IV_all} shows I-V uniformity dependency. The measurement is carried out at room temperature. The bias voltage is scanned in steps of 5~V, whilst it is 1~V scanning step close to breakdown voltage. As shown in Figure~\ref{fig:IV_all}-(a), the leakage current shifts a little according to the position of the sensor on the wafer due to the non-uniformity of doping. As shown in Figure~\ref{fig:IV_all}-(b), the I-V curves are measured with two probe needles on different position of the GR and the DC ring, as marked in Figure~\ref{fig:sensor}. The obtained I-V curves are all the same which indicates that the $n^+$ doping within the active region of the $1\times 1~\mathrm{mm^2}$ sensor can be considered as being uniform.

%------------------------------------------------------------
\subsection{Capacitance-voltage measurement}
\label{sec:C-V test}

The capacitance-voltage (C-V) curves indicate performance of multiplication layer inside the AC-LGAD sensors. The AC-LGAD sensors are tested at room temperature, and bias voltage is scanned in steps of 1~V.

\begin{figure}[htbp]
\centering 
\subfloat[C-V curves.]{\includegraphics[width=.45\textwidth]{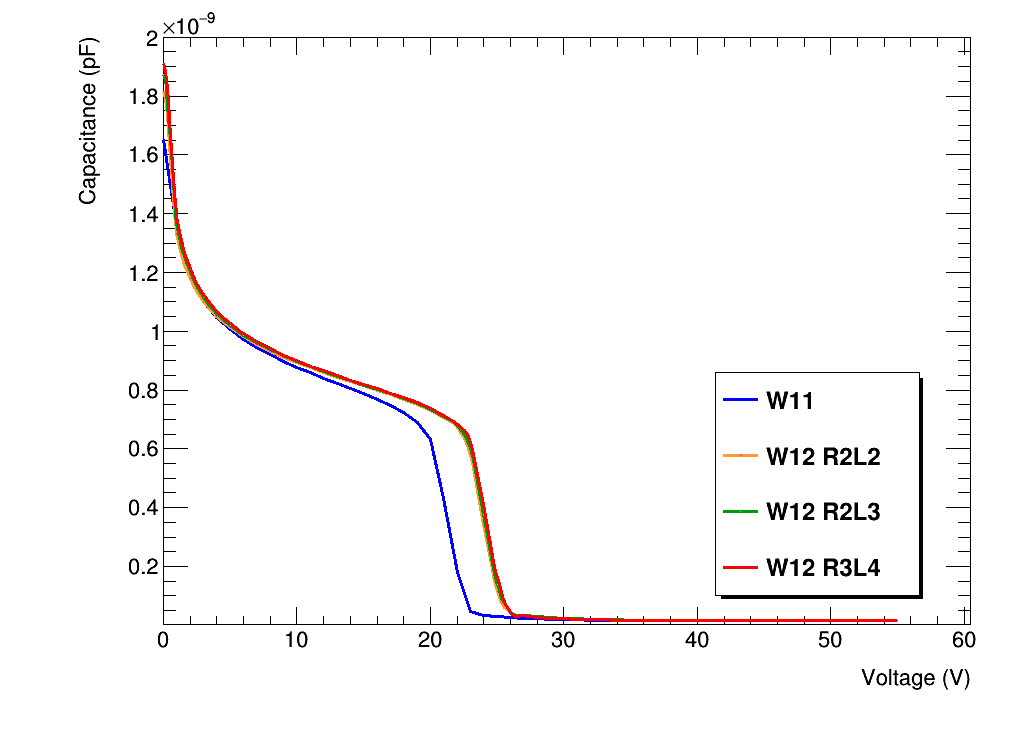}}
\qquad
\subfloat[$\mathrm{1/C^2}$ curves.]{\includegraphics[width=.45\textwidth]{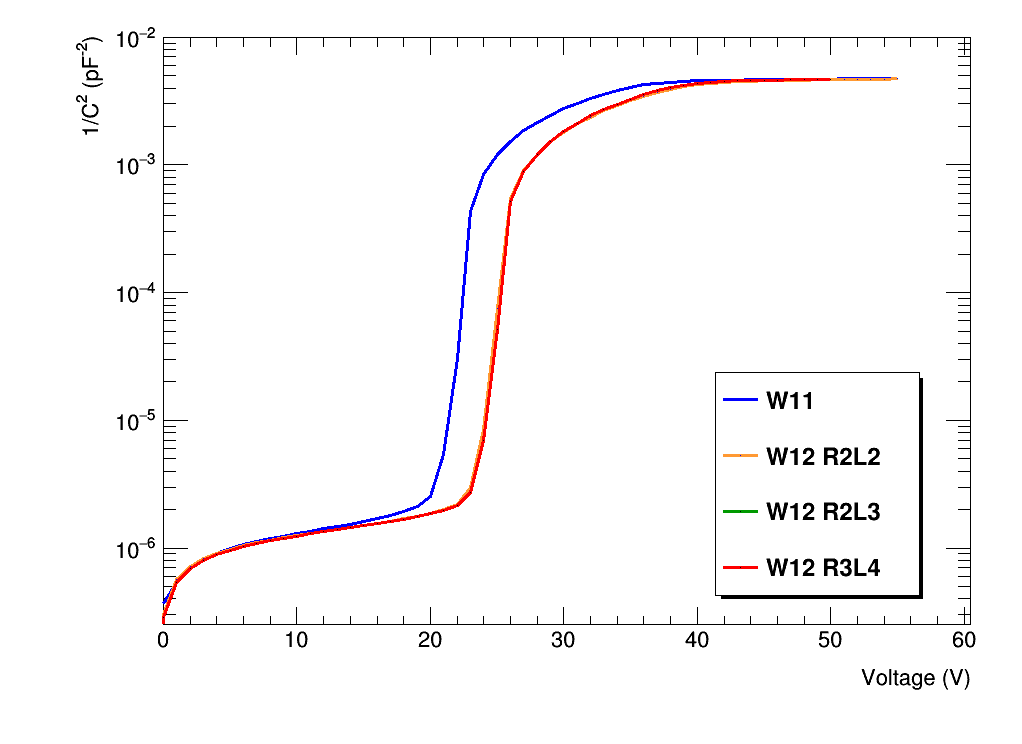}}
\caption{\label{fig:CV} Capacitance-Voltage and $\mathrm{1/C^2}$ curves of the prototype strip sensors are measured at room temperature. The curve of one wafer-11 sensor is shown in blue. The curves of sensors from different position on wafer-12 are shown in yellow, green and red respectively.}
\end{figure}

Figure~\ref{fig:CV}-(a) shows the measured C-V curves of one wafer-11 sensor and three wafer-12 sensors, while figure~\ref{fig:CV}-(b) shows the corresponding $\mathrm{1/C^2}$-V curves. No obvious difference in the C-V curves has seen for sensors from different position on wafer-12. It shows that the doping uniformity among wafer has negligible effect on the junction capacitance. There are two plateaus on the $\mathrm{1/C^2}$ curves. The $\mathrm{1/C^2}$ curves start rising after the first plateau at the voltage of around 20~V (24~V) for wafer-11 (wafer-12) sensors, which voltages indicate that their gain layers are fully depleted. After gain layer depletion, the curves rising part of wafer-11 and wafer-12 sensors are similar, since they have the same bulk resistivity. The wafer-11 and wafer-12 sensors both saturate and enter a second plateau at approximately 40~V voltage value.

%The gain layer depletion $V_{\mathrm{GL}}$ is defined as the voltage where the $\mathrm{1/C^2}$ curves start rising after the first plateau. It is find to be around 20~V (24~V) for the wafer-11 (wafer-12) sensors. 

%The gain layer depletion voltage ($V_{\mathrm{GL}}$) and the full-depletion voltage ($V_{\mathrm{FD}}$) can be obtained through the $\mathrm{1/C^2}$-V curves. $V_{\mathrm{GL}}$ is defined as the voltage where the $\mathrm{1/C^2}$ curves start rising after the first plateau. It is correlated with the maximum of the $n^+$ doping density, and is find to be around 20~V (24~V) for the wafer-11 (wafer-12) sensors. $V_{\mathrm{FD}}$ is defined as the voltage where the curves finally saturate and enter the second plateau\cite{In-depthStudy}. After gain layer depletion, the rising part of the curves of wafer-11 and wafer-12 sensors are similar, since they have the same bulk resistivity. The $V_{\mathrm{FD}}$ is find to be around 40~V for both wafer-11 and wafer-12 sensors.

%------------------------------------------------------------
%------------------------------------------------------------
\section{Position reconstruction and spatial resolution test}
\label{sec:Position reconstruction and spatial resolution}

%------------------------------------------------------------
\subsection{Measurement setup}
\label{sec:Measurement setup 2}

\begin{figure}[htbp]
\centering 
\subfloat[Sensor bonded to a readout board.]{\includegraphics[width=.45\textwidth]{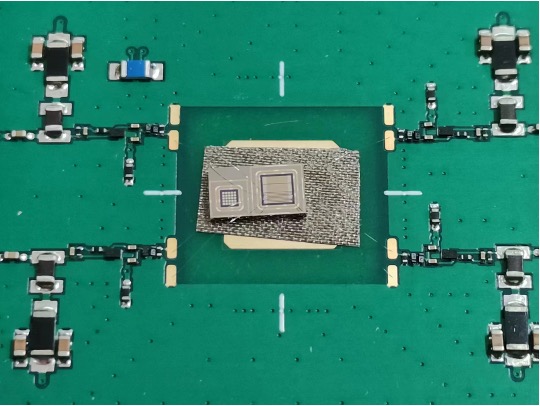}}
\qquad
\subfloat[TCT platform.]{\includegraphics[width=.45\textwidth]{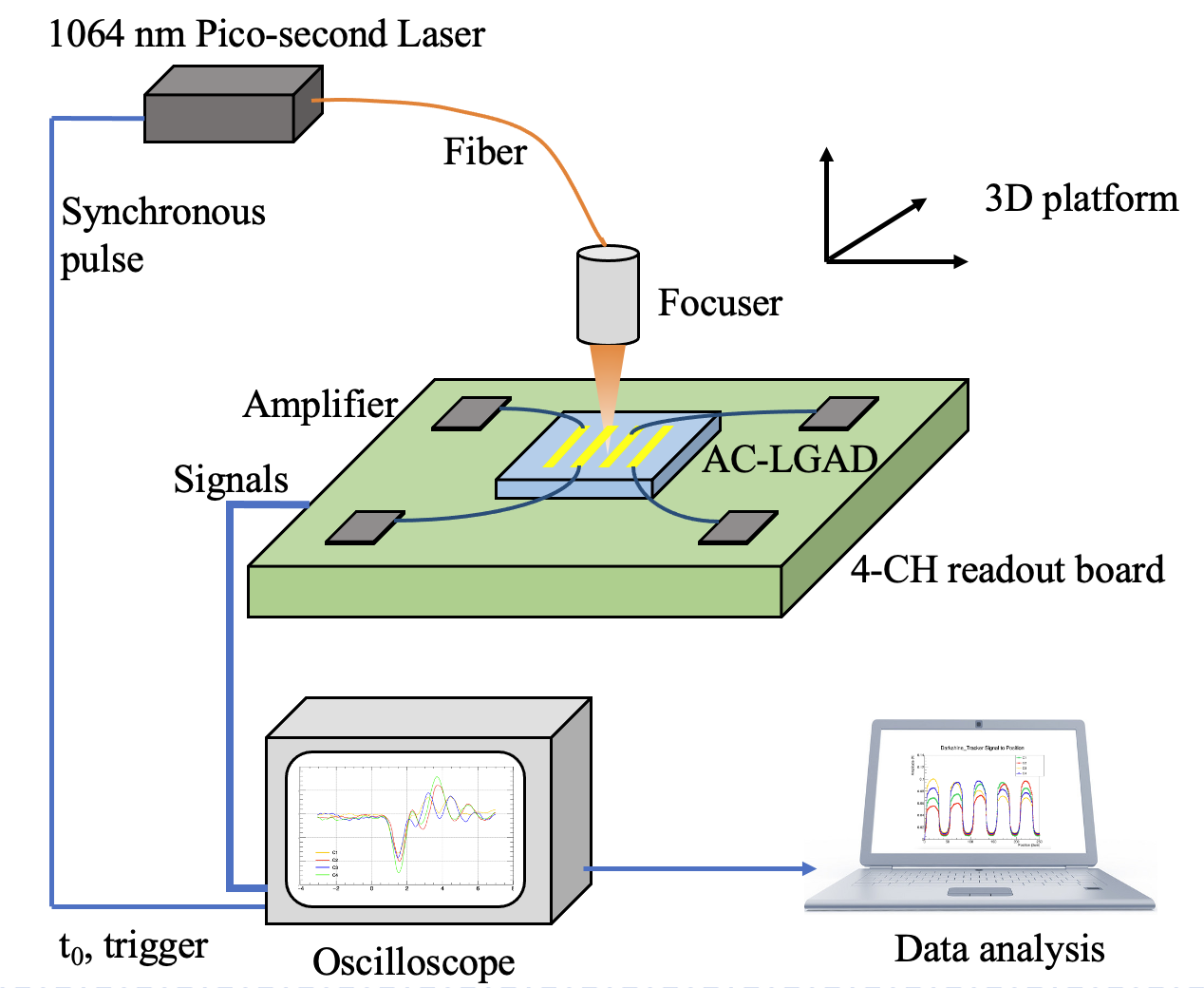}}
\caption{\label{fig:LaserSetup} (a) Picture of sensor bonded to a 4-channel readout PCB board using aluminium wire; (b) Schematic diagram of the laser TCT platform.}
\end{figure}

To investigate spatial resolution of the AC-LGAD sensor, four adjacent strips on the sensor are tested on a transient current technique (TCT) platform which are bonded to a 4-channel readout PCB board. As shown in Figure~\ref{fig:LaserSetup}-(a), a 4-channel readout PCB board is developed and is produced at IHEP with reference to the single-channel readout board designed by University of California Sanata Cruz\cite{CARTIGLIA201783}. A bias voltage is supplied through the back of sensors while its guard ring is grounded. The working voltage is set to $-350$~V ($-150$~V) for wafer-11 (wafer-12) sensor according to its I-V curve measured in Section~\ref{sec:I-V test}. Only 100~$\mathrm{\mu m}$ pitch design is tested in this paper.

The setup of TCT platform is illustrated in Figure~\ref{fig:LaserSetup}-(b). The laser wavelength is 1064~nm. 
%The signal pulse width produced by the laser is 7.68~ps with a frequency of 21.9~MHz. 
The laser spot, with size of about 6~$\mathrm{\mu m}$, can be moved along either $x$ or $y$ direction which is controlled by a 3-dimensional translation platform with accuracy around 1~$\mathrm{\mu m}$. In order to simulate single photon event, an attenuator is placed above the sensor.
%which reduces the laser beam to $0.32\%$ of its original energy. 
The signal is triggered by a synchronous pulse from laser with a potential trigger time shift of around 15~ps. After amplification the signal pulses readout from four wire bonded strips are recorded by digital oscilloscope. It has a sampling width of 20~Gs/s and bandwidth of 1~GHz.

\begin{figure}[htbp]
\centering 
\subfloat[Sensor strips.]{\includegraphics[width=.45\textwidth]{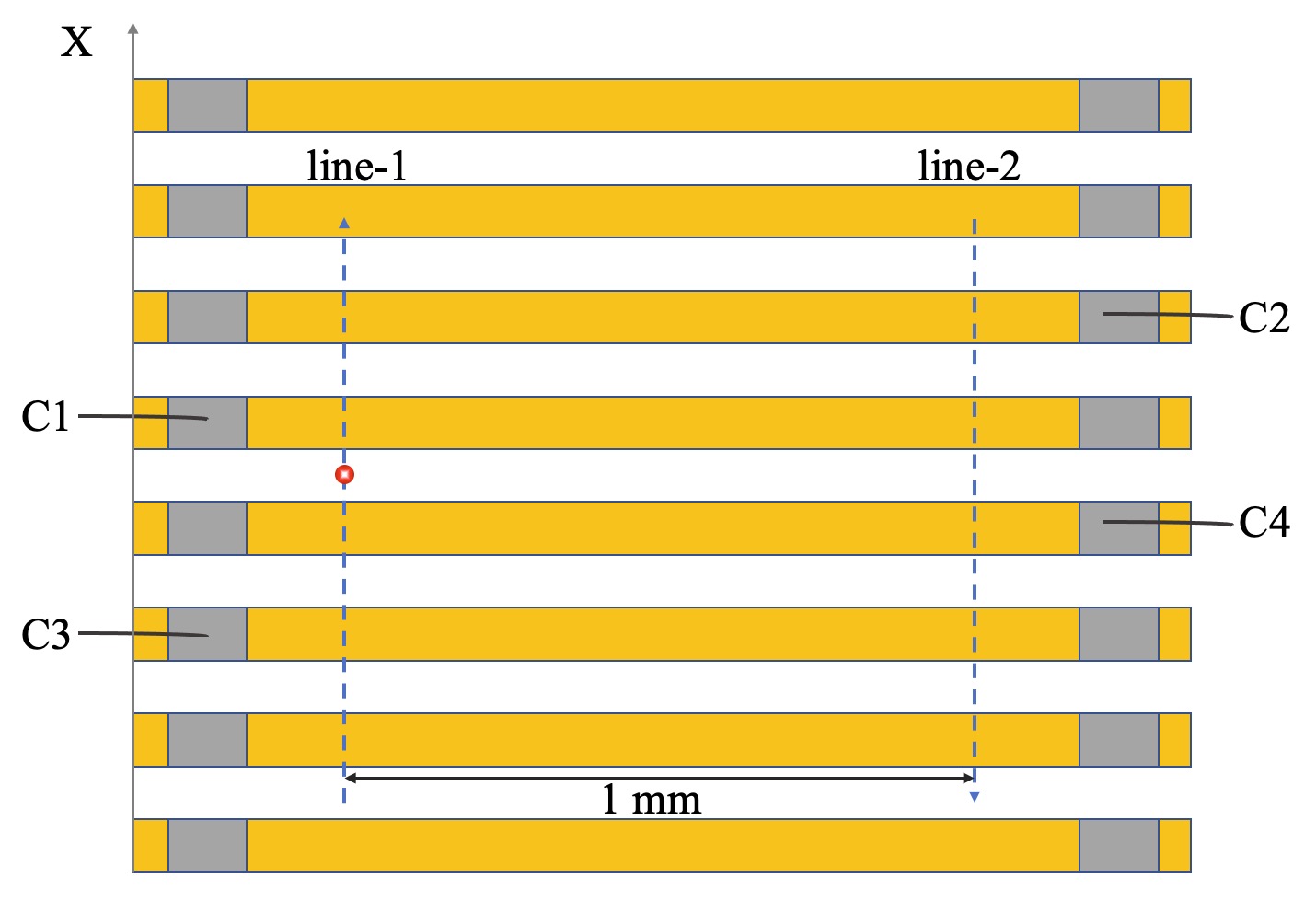}}
\qquad
\subfloat[Signal waveforms.]{\includegraphics[width=.45\textwidth]{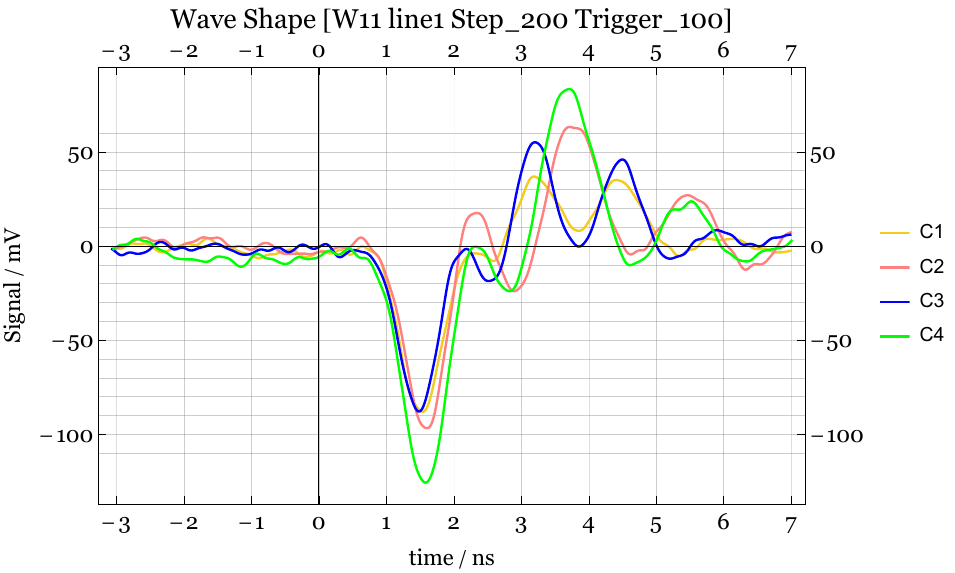}}
\caption{\label{fig:procedure} (a) Schematic diagram of sensor strips. Two lines along the $x$ coordinate indicate the paths of laser spot moving. C1 to C4 represent four strips wire bonded for signal readout. (b) Waveforms of one triggered event. The yellow, red, blue and green curves represent four readout channels corresponding to C1, C2, C3 and C4.}
\end{figure}

 The laser spot is moved with 2~$\mathrm{\mu m}$ step size along the $x$ plus and minus directions perpendicular to the strips, which is illustrated as line-1 and line-2 in Figure~\ref{fig:procedure}-(a). The $y$ coordinate is fixed during the measurement. 
 During a data-taking period the oscilloscope is triggered by the laser around 1300 times. 
 Four signal pulses are recorded in each triggered event which is shown as the deepest peaks in Figure~\ref{fig:procedure}-(b). The signal waveforms have typically 1~ns width. The dependence of peak amplitudes comes from different distances between the readout strips and the laser spot. The peaks with opposite polarity come from electron diffusion within $n^+$ layer towards the edge of the sensor\cite{Giacomini:2019kqz}.

%The distance between line-1 and line-2 is set to be 1~mm, so that we can check the impact from the $y$ position of the laser spot on the signal waveform.

 %For each step, the known $x$ coordinate of the laser spot is taken as the reference for position reconstruction.

%due to the incompatible capacitance of the sensor and the readout board. 
%The latter charge movement induce signals also in nearby readouts. This effect can be attenuated by making the n+ layer to float through a large value resistor connected between ground and the n++ contact.
%No obvious difference of the signal pulse is seen for the same $x$ position on line-1 and line-2.

%------------------------------------------------------------
\subsection{Position reconstruction and spatial resolution}
\label{sec:Position reconstruction}
%The position of the incoming signal (i.e., the laser spot centre) is reconstructed using the amplitude imbalance among the four adjacent readout strips. We take the line-2 measurement of the wafer-11 sensor as an example. Figure~\ref{fig:amp_vs_position_and_slope}-(a) shows the maximum amplitudes of the four channels, averaged over all triggers in each step. The signal amplitude decreases with the distance between the readout strip and the laser spot. However, the signal is not linearly attenuated, thus we need to take a further look into its change rate, as shown in Figure~\ref{fig:amp_vs_position_and_slope}-(b). When the laser spot is blocked by the metal strips, the collected charge or signal amplitude reduces to zero, and the change rate reaches a local maximum or minimum at the edge of the metal strips. From Figure~\ref{fig:amp_vs_position_and_slope}-(c), it can be tell that the signal amplitudes of the most adjacent strips change faster than the next-to-adjacent ones with respect to the position of laser spot. Using these information, the serial number of steps can be translated into $x$ coordinates, and the four ``neighbor strips'' of each step can be found. As shown in Figure~\ref{fig:amp_vs_position_and_slope}-(d), we set $x = 0$ at the centre of the gap between C1 and C4, then extract the signal pulse information form the steps inside this gap for further data analysis.
A simplified linear model\cite{NIMPRA:2021AT} is used to reconstruct the position of signal (i.e. the laser spot centre). This linear model assumes that signal on each pad decreases linearly with its distance from the point of particle incidence. For strip sensors, the position information recorded by four strip sensors has only one dimension. As shown in figure~\ref{fig:procedure}, signal fraction is defined as ratio of the lowest peak of each channel to the lowest peak of all channels. It can be assumed that the signal fraction on each strip varies linearly with the distance from the impact point. The relationship between the impact position and the signal fraction of each strip can be expressed by $$
x=(f_i-\alpha_i)/\beta_i, i=1,2,3,4
$$
where $x$ is the impact position, $f_i$ is the signal fraction of each channel, $\alpha_i$ is the signal fraction of each channel at $x=0$, and $\beta_i$ is the change rate of the signal fraction of each channel with the impact position.

\begin{figure*}[t]
\centering
\subfloat[]{
\includegraphics[width=.47\textwidth]{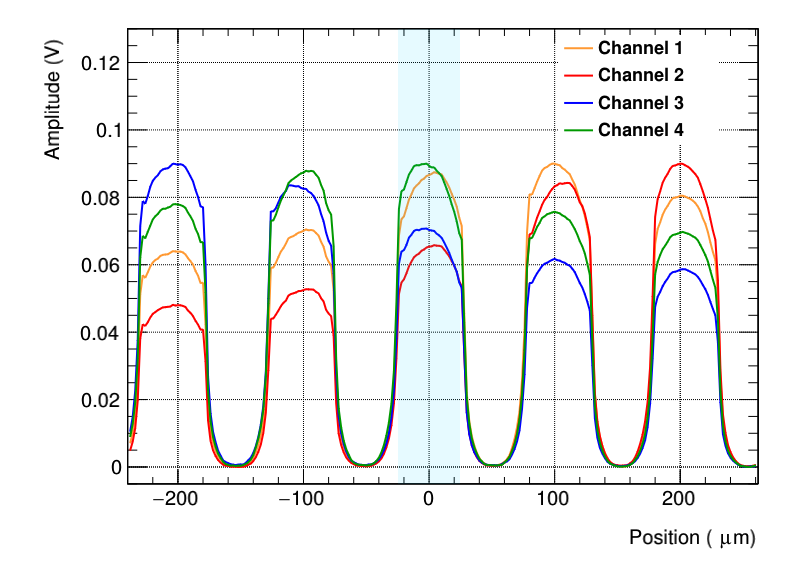}
}
\subfloat[]{
  \includegraphics[width=.45\textwidth]{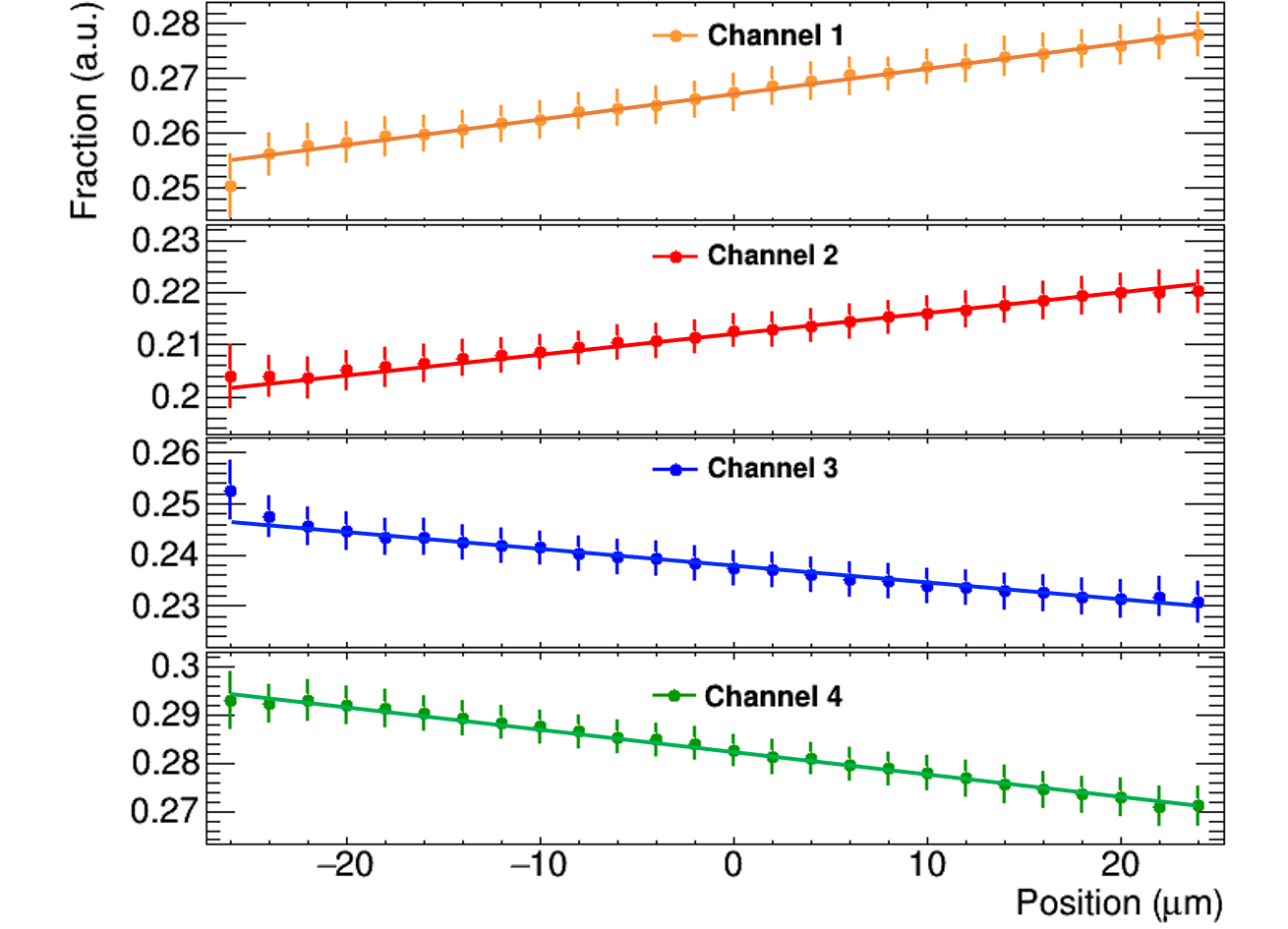}
}\\
%\subfloat[]{
%\includegraphics[width=.47\textwidth]%{Plots/Analysis/Channel_3.png}
%}
%\subfloat[]{
%  \includegraphics[width=.47\textwidth]{Plots/Analysis/Channel_4.png}
%}
\caption{
%(a) The average maximum amplitudes with respect to the $x$ coordinate, while $x = 0$ is set to be at the centre of the gap where the laser hits. The absolute value of amplitudes after normalization are shown in the plot. 
%(b) The change rate of the maximum amplitude with respective to $x$.
%(c) The average maximum amplitudes with respect to the $x$ coordinate, zooming in the selected 25 steps corresponding to the central gap. 
%(d) The change rate of the maximum amplitude with respective to $x$, zooming in the selected 25 steps corresponding to the central gap. Linear fits (colored lines) are applied to the change rate in each channel (colored dots). 
(a) The average maximum amplitudes with respect to the $x$ coordinate, while $x = 0$ is set to be at the centre of the gap where the laser hits. The absolute value of amplitudes after normalization are shown in the plot. Highlighted areas are selected for data analysis.
All plots are measured with line-1 of the wafer-12 sensor. 
(b) Fraction of the maximum amplitude of each readout channel, averaged over all triggered events. Linear fits are applied to each fraction, as shown in colored dashed lines. The yellow, red, blue and green curves represent the four readout channels C1, C2, C3 and C4 of the oscilloscope respectively.}
\label{fig:amp_vs_position_and_slope}
\end{figure*}

Figure~\ref{fig:amp_vs_position_and_slope}-(a) shows the maximum amplitudes of the four channels, which is averaged over all triggers in each step. 
%The signal fraction of each channel is shown at the bottom of each corresponding graph. 
When laser spot is blocked by the metal strips, the collected charge or signal amplitude reduces to zero. We set $x = 0$ at centre of the gap between C1 and C4. This is derived by finding the edges of the metal strip after deriving the signal distribution for each channel.

Figure~\ref{fig:amp_vs_position_and_slope}-(b) shows the fraction of the maximum amplitude of each readout channel as a function of $x$, computed by $$
f_i=\frac{A^i_{max}}{A^1_{max}+A^2_{max}+A^3_{max}+A^4_{max}}, i=1, 2, 3, 4
$$
where $A^i_{max}$ is the maximum amplitude of each channel at given $x$. A linear function is then used to fit the amplitude fraction of each channel. Therefore, four $x$ values can be solved from the fit function for any given event with measured amplitude fraction $(f_1, f_2, f_3, f_4)$, and the average among $(x_1, x_2, x_3, x_4)$ is taken as the reconstructed laser spot position. 

\begin{figure}[htbp]
\centering 
\subfloat[]{\includegraphics[width=.42\textwidth]{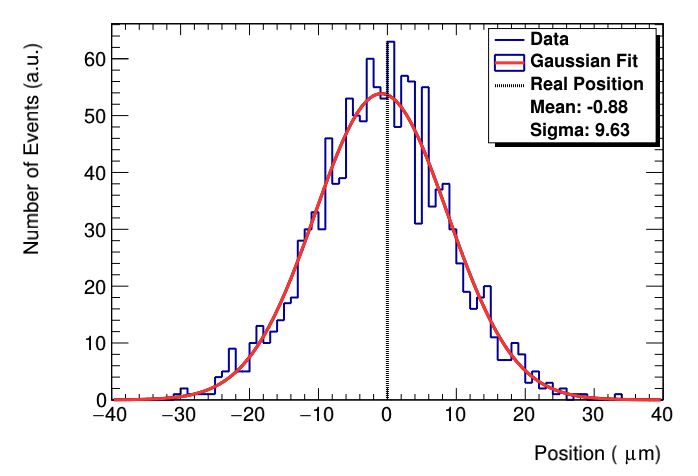}}
\qquad
\subfloat[]{\includegraphics[width=.47\textwidth]{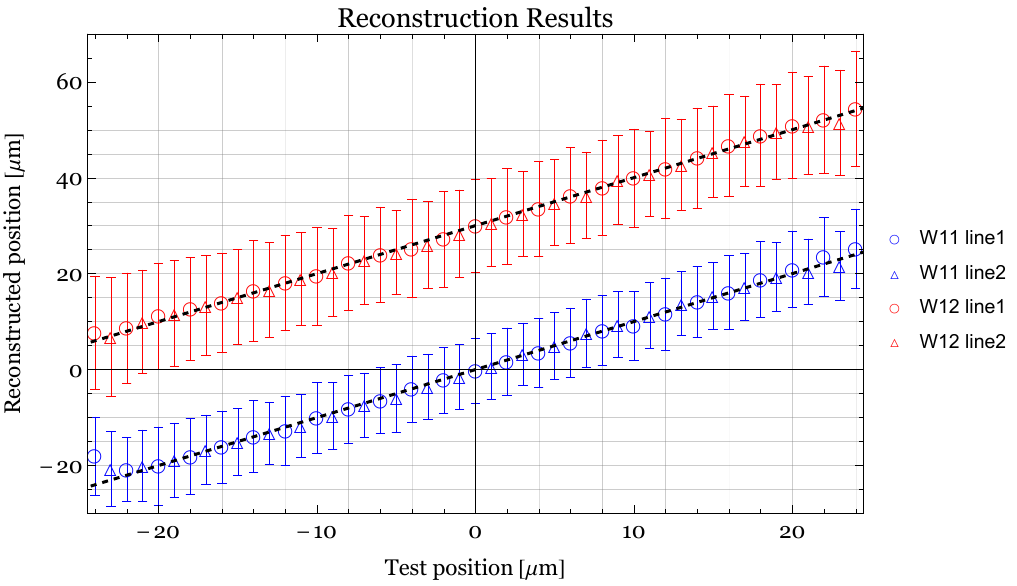}}
\caption{\label{fig:Ampfraction} (a) Distribution of the reconstructed $x$ coordinate for actual laser spot position at $x=0~\mu m$. The red curve shows a Gaussian fit to the distribution. (b) Reconstruction results of the two sensors from wafer-11 and wafer-12. The dashed line ($y=x$) represents the actual laser spot position. The results of wafer-12 sensor are shifted up by 30 $\mathrm{\mu m}$ for better demonstration.}
\end{figure}

Figure~\ref{fig:Ampfraction}-(a) shows a distribution of the reconstructed $x$ coordinates for a known test position at $x=0~\mu m$. From a Gaussian fit to this distribution, its mean value -0.88 $\mu m$ is taken as the reconstructed $x$, and its sigma value 9.63 $\mu m$ is considered as the spatial resolution.

\begin{figure}[htbp]
\centering 
\subfloat[Spatial resolution.]{
  \includegraphics[width=.47\textwidth]{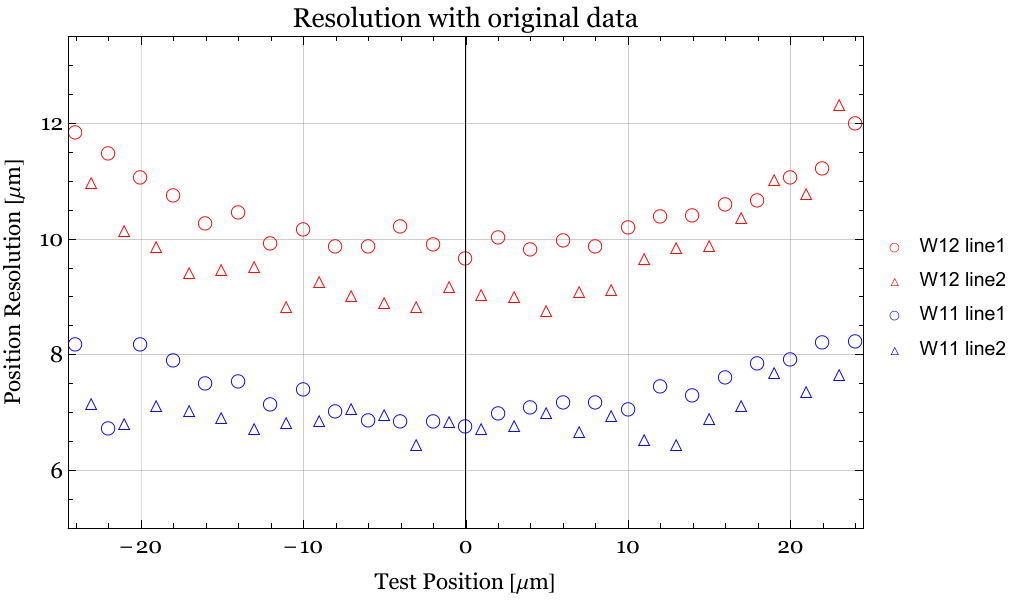}
}
\subfloat[standard deviation of signal fraction.]{
  \includegraphics[width=.44\textwidth]{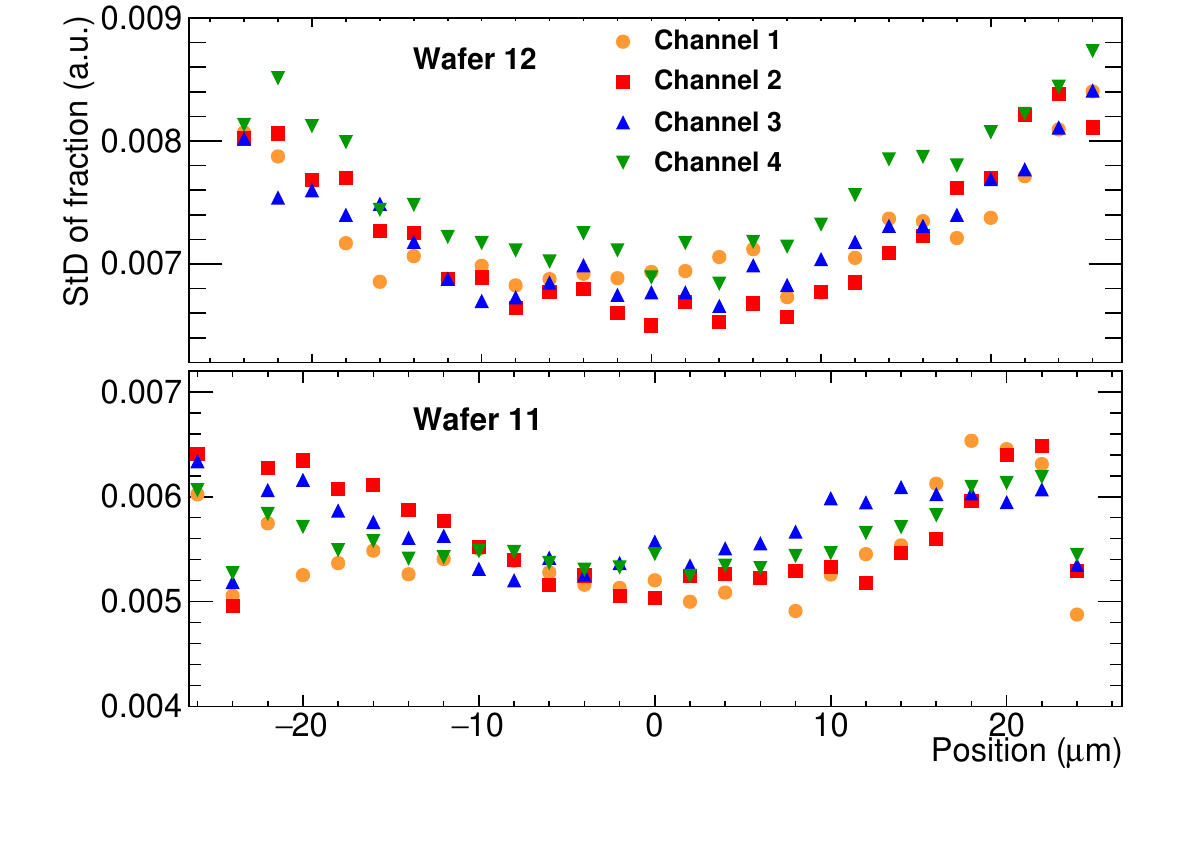}
}  
\caption{\label{fig:results} (a) Distribution of spatial resolution with respect to $x$ coordinates. The blue and red marks represent the reconstructed position for wafer-11 and wafer-12 sensor, respectively. The circle represents measurement from line-1, while triangle represents measurement from line-2. (b) The distribution of the standard deviation of the signal fraction with respect to the $x$ coordinate}
\end{figure}

Figure~\ref{fig:Ampfraction}-(b) shows two sets of reconstructed positions from wafer-11 and wafer-12 and a comparison with spot positions. The dashed lines indicate spot positions while the dots are reconstructed positions. They are in good agreement. Figure~\ref{fig:results}-(a) shows the corresponding spatial resolution as a function of $x$, which varies from 6.5~$\mathrm{\mu m}$ to 8.2~$\mathrm{\mu m}$ for wafer-11 and from 8.8~$\mathrm{\mu m}$ to 12.3~$\mathrm{\mu m}$ for wafer-12 respectively. Better resolution of wafer-11 sensor than wafer-12 is due to the fact that $n^+$ dose (0.01~P) of wafer-11 is much smaller than wafer-12 (10~P). The spatial resolution is slightly worse in the edge of gaps as shown in figure~\ref{fig:results}-(b). It is due to the fact that signal fraction is less stable when laser spot closing to metal strips. Similar effect also leads to a small overall discrepancy between the line-1 and line-2 measurements. 
%It might come from the different laser spot size in the two measurements, resulting in slightly larger variance of line-1. 
%Performance of the sensor from the both two wafers satisfy the requirement from the designed parameters of the DarkSHINE tracking system.

%------------------------------------------------------------
%------------------------------------------------------------
\section{Summary}
\label{sec:Summary}

This paper presents the performance of two batches of prototype AC-LGAD sensors with pitch size of 100~$\mu m$ designed for the DarkSHINE experiment. The range of spatial resolutions are 6.5~$\mathrm{\mu m}$ $\sim$ 8.2~$\mathrm{\mu m}$ and 8.8~$\mathrm{\mu m}$ $\sim$ 12.3~$\mathrm{\mu m}$ for wafer-11 and wafer-12 sensors. The typical sensor response time is about 1~ns. The wafer-11 sensor delivers better spatial resolution because of smaller $n^+$ dose. Both wafer-11 and wafer-12 sensors satisfy the requirement of the DarkSHINE tracking system.

This work has been supported by a grant from the National Natural Science Foundation of China (Grant NO.12150006).
%------------------------------------------------------------
%------------------------------------------------------------



\begin{thebibliography}{99}

\bibitem{DarkSHINE}
J. Chen, et al, Prospective study of light dark matter search with a newly proposed DarkSHINE experiment, \emph{Sci. China Phys. Mech. Astron.}  {\bf 66} (2023) 211062.

\bibitem{darkphoton}
M. Fabbrichesi, E. Gabrielli, and G. Lanfranchi, The dark photon. arXiv: 2005.01515

\bibitem{HFW.Sadrozinski}
 H.F.W. Sadrozinski et al., Ultra-fast silicon detectors, \emph{Nucl. Instrum. Meth.} \textbf{A 730} (2013) 226.

\bibitem{ATLAS}
ATLAS collaboration, Technical Proposal: A High-Granularity Timing Detector for the ATLAS Phase-II Upgrade. CERN-LHCC-2018-023.

\bibitem{CMS}
CMS collaboration, Technical proposal for a MIP timing detector in the CMS experiment phase 2 upgrade, CERN-LHCC-2017-027.

\bibitem{Giacomini:2019kqz}
G. Giacomini, W. Chen, G. D’Amen and A. Tricoli, Fabrication and performance of AC-coupled LGADs, \emph{JINST} {\bf 14} (2019) P09004.

\bibitem{N+dosePaper}
M. Li, et al., The performance of large-pitch AC-LGAD with different N+ dose, (2022) arxiv:2212.03754.

\bibitem{RD50Recommendation}
A. Chilingarov, RD50-Recommendations for performing measurements. Part I: IV and CV measurements in Si diodes.

\bibitem{CARTIGLIA201783}
N. Cartiglia, et al., Beam test results of a 16 ps timing system based on ultra-fast silicon detectors, \emph{Nucl. Instrum. Methods Phys. Res. A} {\bf 850} (2017) 83-88.

\bibitem{NIMPRA:2021AT}
M.Tornago, R. Arcidiacono et al., Resistive AC-Coupled Silicon Detectors: \emph{Nucl. Instrum. Methods Phys. Res. A} 1003(2021) 165319.

%----something for SHINE----------
%\bibitem{Wan:2022het}
%J.~Wan, Y.~Leng, B.~Gao, F.~Chen, J.~Chen, L.~Lai, W.~Zhou and W.~Xu,
%``Simulation of wire scanner for high repetition free electron laser facilities,''
%\textit{Simulation of wire scanner for high repetition free electron laser facilities},
%Nucl. Instrum. Meth. A \textbf{1026}, 166200 (2022)
%Nucl. Instrum. Meth. A 1026, 166200 (2022)
%doi:10.1016/j.nima.2021.166200
%0 citations counted in INSPIRE as of 30 Mar 2022 

%\cite{Zhao:2017ood}
%\bibitem{Zhao:2017ood}
%Z.~T.~Zhao, C.~Feng and K.~Q.~Zhang,
%``Two-stage EEHG for coherent hard X-ray generation based on a superconducting linac,''
%\textit{Two-stage EEHG for coherent hard X-ray generation based on a superconducting linac},
%Nucl. Sci. Tech. \textbf{28}, no.8, 117 (2017)
%Nucl. Sci. Tech. 28, no.8, 117 (2017)
%doi:10.1007/s41365-017-0258-z
%5 citations counted in INSPIRE as of 14 May 2022

%\cite{Zhao:2016ood}
%\bibitem{Zhao:2016ood}
%Z.~T.~Zhao, C.~Feng, J.~H.~Chen and Z.~Wang,
%``Two-beam based two-stage EEHG-FEL for coherent hard X-ray generation,''
%\textit{Two-beam based two-stage EEHG-FEL for coherent hard X-ray generation},
%Science Bulletin \textbf{61}, 117 (2016), 720-727
%Science Bulletin 61, 117 (2016), 720-727
%doi:10.1007/s11434-016-1060-8

%\cite{Zhao:2018lcl}
%\bibitem{Zhao:2018lcl}
%Z.~Zhao, D.~Wang, Z.~H.~Yang and L.~Yin,
%``SCLF: An 8-GeV CW SCRF Linac-Based X-Ray FEL Facility in Shanghai,''
%doi:10.18429/JACoW-FEL2017-MOP055
%24 citations counted in INSPIRE as of 16 Jun 2022

%\cite{Nosochkov:2017xoc}
%\bibitem{Nosochkov:2017xoc}
%Y.~Nosochkov, T.~Beukers, A.~Fry, C.~Hast, T.~Markiewicz, T.~Nelson, N.~Phinney, T.~Raubenheimer, P.~Schuster and N.~Toro,
%``Dark Sector Experiments at LCLS-II (DASEL) Accelerator Design,''
%doi:10.18429/JACoW-IPAC2017-TUPIK121
%0 citations counted in INSPIRE as of 14 Jun 2022
%----something for SHINE----------


%\bibitem{ATLAS}
%G. Aad et al, The ATLAS Experiment at the CERN Large Hadron Collider. \emph{JINST} \textbf{3} (2008), S08003.

%\bibitem{CMS}
%S. Chatrchyan et al, The CMS Experiment at the CERN LHC. \emph{JINST} \textbf{3} (2008), S08004.



%\bibitem{dopants}
%L. Rossi, P. Fischer, T. Rohe, N. Wermes, Pixel Detectors: From Fundamentals to Applications, \emph{Springer Science & Business Media} (2006).

%\bibitem{In-depthStudy}
%H. Sadrozinski, In-depth experimental study of acceptor removal in low-gain avalanche detectors, in: \emph{12th International Hiroshima Symposium on Development and Application of Semiconductor Tracking Detectors, UCSC} (2019)

%\bibitem{Keithley}
%Dual-channel picoammeter/voltage source 6482 datasheet.


% Please avoid comments such as "For a review'', "For some examples",
% "and references therein" or move them in the text. In general,
% please leave only references in the bibliography and move all
% accessory text in footnotes.

% Also, please have only one work for each \bibitem.


\end{thebibliography}
\end{document}